\documentclass[showpacs,twocolumn,showkeys,amsmath,amssymb,superscriptaddress,aps,prl]{revtex4}
\usepackage{amssymb} \usepackage{amsmath} %\usepackage{caption2}
\usepackage{dcolumn}% Align table columns on decimal point
\usepackage{bm}% bold math \include{natbib} %\nofiles

\begin{document}
\title{Tsallis Ensemble as an Exact Orthode}
%\title{Tsallis Ensemble as a mechanical Model of nonextensiveThermodynamics}
\author{M. Campisi}\thanks{Corresponding
author} \email{campisi@crim.sssup.it} \affiliation{CRIM Lab,
Scuola Superiore Sant'Anna, Viale R. Piaggio 34, 56025 Pontedera
(PI) Italy} \author{G. B. Bagci} \affiliation{Department of
Physics, University of North Texas, P.O. Box 311427, Denton, TX
76203-1427, USA} \date{\today}

\begin{abstract} We show that
Tsallis ensemble of power-law distributions provides a mechanical
model of nonextensive equilibrium thermodynamics for small
interacting Hamiltonian systems, i.e., using Boltzmann's original
nomenclature, we prove that it is an exact orthode. This means
that the heat differential admits the inverse average kinetic
energy as an integrating factor. One immediate consequence is that
the logarithm of the normalization function can be identified with
the entropy, instead of the q-deformed logarithm. It has been
noted that such entropy coincides with R\'{e}nyi entropy rather
than Tsallis entropy, it is non-additive, tends to the standard
canonical entropy as the power index tends to infinity and is
consistent with the free energy formula proposed in [S. Abe
\emph{et. al.} Phys. Lett. A {\bf 281}, 126 (2001)]. It is also
shown that the heat differential admits the Lagrange multiplier
used in non-extensive thermodynamics as an integrating factor too,
and that the associated entropy is given by ordinary nonextensive
entropy. The mechanical approach proposed in this work is fully
consistent with an information-theoretic approach based on the
maximization of R\'{e}nyi entropy.
\end{abstract}

\maketitle After the pioneering work of Tsallis \cite{Tsallis88},
nonextensive thermostatistics has been studied and applied to many
diverse fields such as quantum information \cite{Abe04} and high
energy collisions \cite{Wilk00}. Such a great success is mainly
due to the fact that non-extensive thermostatistics proved to be a
successful tool for studying a wide variety of complex phenomena,
ranging from turbulence \cite{Bogho96} to astrophysics
\cite{Plast93}, which exhibit anomalous behavior characterized by
self-similarity and the emergence of power-law distributed events.
Despite its success, non-extensive thermostatistics is not yet
completely well founded. For example, important problems such as
the definition of physical temperature of a system obeying Tsallis
statistics \cite{Abe01}, and whether it is necessary to use
ordinary or normalized expectation values, are still under debate
\cite{Abe05}. Historically, in the literature of non-extensive
thermostatistics, the Tsallis ensemble of distributions is derived
from the maximization of a generalized form of Shannon entropy
(namely the so-called Tsallis entropy), subject to the constraints
of the normalization and internal energy. In this Letter, we shall
address the problem of the foundation of Tsallis statistics from a
different point of view whose roots can be traced back to
Boltzmann himself. The method, recently reviewed by Gallavotti
\cite{Gallavotti}, consists in checking whether the ensemble is an
\emph{orthode}, i.e., checking whether the \emph{heat theorem},
\begin{equation}\label{eq:heat-theo}
    \frac{\delta Q}{T} = exact\quad differential
\end{equation}
holds within the ensemble. The concept of \emph{orthodicity} was
developed by Boltzmann who placed it at the very foundation of the
theory of ensembles in statistical mechanics \cite{Gallavotti}.
Unfortunately, this method seems not to have gained the same
popularity like some other methods based on counting and
information theory (also traceable back to the prolific scientific
creativity of Boltzmann). Few important exceptions to this trend
are represented by a couple of textbooks concerned with the
mathematical foundations of statistical mechanics
\cite{Gallavotti,Khinchin}.

The main contribution of this work is represented by the
acknowledgement that Tsallis ensemble is an orthode (indeed an
exact orthode) namely the integrating factor $(1/T)$ exists and is
equal to the average kinetic energy per degree of freedom, thus
being in complete agreement with what happens within canonical,
microcanonical and other standard ensembles. Quite surprisingly,
though, the corresponding entropy is not given by Tsallis formula,
but, as we shall see, a connection exists between the present
approach and the standard nonextensive treatment. Questions
concerning the heat theorem within the non-extensive framework
have been addressed only very recently in Refs. \cite{Ponno06} and
\cite{Abe06}. The present contribution differs from that of Ref.
\cite{Ponno06} in the sense that the Boltzmanian point of view has
been adopted rather than the Gibbsian one in this Letter. The
other main difference concerns the form of the distribution
studied, which, in this Letter is implicitly defined in terms of
the average energy $E$ (see Eq. (\ref{eq:Tsallis-ensemble})
below), as it results from the Tsallis entropy maximization
procedure \cite{Abe05}. In this respect the present work is much
closer to the work by \citeauthor{Abe06} \cite{Abe06}.

In order to proceed to present the results, let us begin by
reviewing the concept of \emph{orthodicity} following Gallavotti's
reconstruction of Boltzmann's original arguments
\cite{Gallavotti}. Consider a family (ensemble) of distributions
defined on the phase space of some Hamiltonian system,
parameterized by a given number of parameters $\lambda_i$. For
example the microcanonical ensemble is written as
$\varrho(\mathbf{z};E,V)=
\frac{\delta(E-H(\mathbf{z};V))}{\Omega(E,V)}$, and the canonical
ensemble is given by $\varrho(\mathbf{z};\beta,V)= \frac{e^{-\beta
H(\mathbf{z};V)}}{Z(\beta,V)}$. Let $f$ be the number of degrees
of freedom and the Hamiltonian be of the type:
\begin{equation}\label{}
    H(\textbf{z};V) =\frac{\textbf{p}^2}{2m}+\varphi(\textbf{q};V),
\end{equation}
where
$\textbf{z}=(\textbf{q},\textbf{p})=(q_1,q_2,..q_f,p_1,p_2,...p_f)$
and $V$ is an external parameter (for example the coordinate of
some other body which interacts with the system, like that of a
movable piston which performs work on gas contained in a vessel).
To any external parameter, which can sometimes be named as
``generalized displacement'', there corresponds a ``generalized
conjugated force''. For example, the pressure is taken to be the
generalized conjugated force corresponding to the generalized
displacement called volume in particular. Here we shall assume,
without loss of generality, that there is only one external
parameter. Let us then define the macroscopic state of the system
by the set of following quantities:
\begin{equation}
\begin{tabular}{l}
$E= \left\langle H \right\rangle $\quad``energy'' \\
$T= \frac{2\left\langle K\right\rangle}{f}$\quad ``doubled kinetic
energy per degree of freedom''\\ $V= \left\langle V \right\rangle
$ \quad ``generalized displacement'' \\ $P=\left\langle
-\frac{\partial H}{\partial V}\right\rangle$ \quad ``generalized
conjugated force''
\end{tabular} \label{stateDef} \end{equation} where the symbol
$\left\langle \cdot \right\rangle$ denotes the average over the
distribution $\rho(\textbf{z};\lambda_i)$. The ensemble is said to
provide a mechanical model of thermodynamics, i.e., it is an
\emph{orthode}, if, for infinitesimal and independent changes of
the $\lambda_i$'s, the \emph{heat theorem}
\begin{equation}\label{}
    \frac{dE+PdV}{T} = exact\quad differential
\end{equation}
holds. It is known, since the seminal works of Boltzmann, that the
canonical and the microcanonical ensembles are orthodic
\cite{Gallavotti,Campisi05}.

Let us now consider the normalized Tsallis ensemble of power-law
distributions \cite{Abe05}:
\begin{equation}\label{eq:Tsallis-ensemble}
    \rho (\textbf{z};E,V) = \frac{\left[ 1 -
\frac{\beta}{\alpha}(H(\textbf{z};V)-E)\right]^{\alpha-1}}{N(E,V)}
\end{equation}
where
\begin{equation}\label{eq:N}
    N(E,V) = \int d\textbf{z} \left[ 1 -
\frac{\beta}{\alpha}(H(\textbf{z};V)-E)\right]^{\alpha-1}
\end{equation}
is the normalization (partition function), and $\beta$ is such
that $E$ satisfies the equation
\begin{equation}\label{eq:E=<H>}
\left\langle H \right\rangle = E,
\end{equation}
as prescribed by the maximum entropy method and sometimes
explicitly assumed in formal treatises of standard statistical
mechanics for the canonical ensemble \cite{Khinchin}. Note that
Eq. (\ref{eq:E=<H>}) implies that $\beta=\beta(E,V)$. For reasons
of simplicity, we introduced the parameter $\alpha$ which is
related to the nonextensivity parameter $q$ via the following
relation $\alpha = \frac{1}{1-q}$. The limit $q \rightarrow 1$
corresponds to the limit $\alpha \rightarrow \infty$. Note that in
Eq. (\ref{eq:Tsallis-ensemble}) we adopted the power law index
$\alpha-1 = \frac{q}{1-q}$, i.e. we are using the escort
distribution to evaluate averages. Let us now introduce the
following quantity:
\begin{equation}\label{eq:EN}
    \mathcal{N}(E,V) = \int d\textbf{z} \left[ 1 -
\frac{\beta}{\alpha}(H(\textbf{z};V)-E)\right]^{\alpha}
\end{equation}
Then it is straightforward to see that the ensemble given by Eq.
(\ref{eq:Tsallis-ensemble}) is an orthode, and the associated
entropy is:
\begin{equation}\label{eq:S}
    S(E,V) = \log \mathcal{N}(E,V),
\end{equation}
namely
\begin{equation}\label{}
dS = {dE+PdV\over T},
\end{equation}
 where $E,V,P,T$
are defined according to Eq. (\ref{stateDef}) as averages over the
Tsallis ensemble (\ref{eq:Tsallis-ensemble}). Before proceeding to
the proof, let us recall two known facts \cite{Martinez02}. First,
Eq. (\ref{eq:E=<H>}) implies that
\begin{equation}\label{eq:n=N}
    \mathcal{N}(E,V) = N(E,V)
\end{equation}
Second, there exists an equipartition theorem associated with the
ensemble (\ref{eq:Tsallis-ensemble}) according to which
\begin{equation}\label{eq:equiTeo}
   2 \left\langle \frac{p_i^2}{2m} \right\rangle = \frac{1}{\beta}
\end{equation}
This can be easily proved via integration by parts, using the
cut-off condition according to which the distribution is null over
the region $1-\beta(H-E)/\alpha < 0$ and using Eq. (\ref{eq:n=N})
\cite{Martinez02}. An immediate consequence of the equipartition
theorem is that:
\begin{equation}\label{eq:T=1/beta}
    T = \frac{1}{\beta}
\end{equation}
In order to prove orthodicity, we write:
%-------------------------------------------------------------------------
 \begin{eqnarray}\label{eq:deENdeE}
    \frac{\partial\mathcal{N}}{\partial E} &=&
    \frac{\partial}{\partial{E}}\int
d\textbf{z} \left[ 1 -
    \frac{\beta}{\alpha}(H-E)\right]^{\alpha} \nonumber \\
    &=& \int d\textbf{z} \left[ 1 -
    \frac{\beta}{\alpha}(H-E)\right]^{\alpha-1}\left[-\frac{\partial
\beta}{\partial E} (H-E) + \beta
    \right]\nonumber \\
    &=& - N \frac{\partial \beta}{\partial E} \langle H-E \rangle +
    N\beta = \mathcal{N} \beta = \frac{\mathcal{N}}{T}
\end{eqnarray}
and
\begin{eqnarray}\label{eq:deENdeV}
    \frac{\partial\mathcal{N}}{\partial V} &=&
\frac{\partial}{\partial{V}}\int d\textbf{z} \left[ 1 -
    \frac{\beta}{\alpha}(H-E)\right]^{\alpha} \nonumber \\
    &=& \int d\textbf{z} \left[ 1 -
    \frac{\beta(H-E)}{\alpha}\right]^{\alpha-1}\left[\frac{\partial
\beta}{\partial V} (E-H) -
    \beta\frac{\partial H}{\partial V}
    \right]\nonumber \\
    &=& N \frac{\partial \beta}{\partial V} \left\langle E-H \right\rangle
    - N \beta \left\langle \frac{\partial H}{\partial V} \right\rangle
     = \mathcal{N} \beta P = \mathcal{N}\frac{P}{T}
\end{eqnarray}
so that:
\begin{equation}\label{eq:dS}
    dS =\frac{d\mathcal{N}}{\mathcal{N}}=\frac{dE+PdV}{T}
\end{equation}
%-------------------------------------------------------------------------
This shows that Tsallis ensemble provides a mechanical model of
thermodynamics \cite{Gallavotti}, in the sense that the
\emph{mechanical} quantities $<H>$, $2<K>/f$, $V$, and $-<\partial
H/
\partial V>$, evaluated over the escort distribution, are related to each
other according to the prescriptions of \emph{thermodynamics}. One
immediate and quite surprising consequence is that $\log
\mathcal{N}$ has to be identified with entropy, instead of Tsallis
entropy.

Quite interestingly the property of orthodicity can be proved also
by adopting a slightly different point of view, namely by
considering the ensemble (\ref{eq:Tsallis-ensemble}), as
parameterized by $(\beta,V)$ rather than $(E,V)$. This means that
for each fixed $(\beta,V)$, $E$ is chosen in such a way as to be
the solution of Eq. (\ref{eq:E=<H>}). Therefore $E$, and
consequently also $N$ and $\mathcal{N}$, are considered as
functions of $(\beta,V)$. In this case one has:
 \begin{eqnarray}\label{eq:deENdeBeta}
    \frac{\partial\mathcal{N}}{\partial \beta} &=&
    \frac{\partial}{\partial{\beta}}\int
d\textbf{z} \left[ 1 -
    \frac{\beta}{\alpha}(H-E)\right]^{\alpha} \nonumber \\
    &=& \int d\textbf{z} \left[ 1 -
    \frac{\beta}{\alpha}(H-E)\right]^{\alpha-1}\left[-(H-E) + \beta\frac{\partial E}{\partial \beta}
    \right]\nonumber \\
    &=& - N \langle H-E \rangle +
    N\beta\frac{\partial E}{\partial \beta} = {\mathcal{N}\over T}\frac{\partial E}{\partial \beta}
\end{eqnarray}
and
\begin{eqnarray}\label{eq:deENdeV2}
    \frac{\partial\mathcal{N}}{\partial V} &=&
\frac{\partial}{\partial{V}}\int d\textbf{z} \left[ 1 -
    \frac{\beta}{\alpha}(H-E)\right]^{\alpha} \nonumber \\
    &=& \int d\textbf{z} \left[ 1 -
    \frac{\beta(H-E)}{\alpha}\right]^{\alpha-1}\left[\beta \frac{\partial
E}{\partial V} - \beta \frac{\partial H}{\partial V}\right]\nonumber \\
    &=& N \beta \frac{\partial
E}{\partial V} - N\beta \left\langle\frac{\partial H}{\partial
V}\right\rangle =  {\mathcal{N} \over T}\left( \frac{\partial
E}{\partial V} + P \right)
\end{eqnarray}
so that:
\begin{equation}\label{}
    dS =\frac{d\mathcal{N}}{\mathcal{N}}=\frac{\frac{\partial E}{\partial \beta}d\beta+ \frac{\partial E}{\partial V}dV + PdV}{T}
\end{equation}
but $dE = {\partial E \over \partial \beta}d\beta+ \frac{\partial
E}{\partial V}dV$. Therefore:
\begin{equation}\label{}
    dS =\frac{dE + PdV}{T}
\end{equation}

This feature can be thought of as a \emph{duality} property which
the Tsallis ensemble (\ref{eq:Tsallis-ensemble}) does not share
with the canonical and microcanonical ensembles. Further it
suggests that the Tsallis ensemble could be thought of as a
\emph{hybrid} ensemble, placed half the way between the canonical
and the microcanonical ensemble.

A few comments on the main result proved above are in order. These
concern few fundamental issues posed by non-extensive
thermodynamics, namely whether canonical results are recovered in
the $\alpha \rightarrow \infty$, whether the ensemble
(\ref{eq:Tsallis-ensemble}), and the corresponding entropy
(\ref{eq:S}) are suitable for describing small dimensional systems
and whether the entropy (\ref{eq:S}) is additive. We shall discuss
this in relation to the standard orthodic ensembles, namely the
microcanonical and the canonical one. Finally we'll establish a
connection between the the non-extensive entropy, based on
information theory, and the present treatment based on the heat
theorem.

Let us begin with the recovery of canonical results. Thanks to the
well known properties of the q-deformed exponential
\cite{Tsallis98}, in the limit $\alpha \rightarrow \infty$, the
expression of entropy given by Eq. (\ref{eq:S}), recovers the
standard entropy of the canonical ensemble \cite{Khinchin}:
\begin{equation}\label{eq:lim-logEN}
    \lim_{\alpha \rightarrow \infty} \log \mathcal{N} = \log \int
    d\textbf{z}
e^{-\beta
    (H-E)}= \beta E + \log \int d\textbf{z} e^{-\beta H}
\end{equation}
Note that in this limit one has $N=\mathcal{N} = e^{\beta E}Z$,
where $Z$ is the canonical partition function. Therefore $N$ is
not really the q-counterpart of $Z$, this is the reason why we
didn't keep the established notation $Z_q$ and introduced the $N$
notation instead.

It is easy to see that if the canonical entropy is additive thanks
to the factorization of the exponential ($e^{-\beta
(H_1+H_2)}=e^{-\beta H_1}e^{-\beta H_2}$ ), the entropy in Eq.
(\ref{eq:S}), is not, because the power-law does not factorize.
Therefore, despite the logarithmic structure, the entropy
(\ref{eq:S}) \emph{is} suitable for describing non-additive
systems. In this respect it is fair to stress a simple fact which
is often overlooked, namely that the entropy associated with the
microcanonical orthode
\begin{equation}\label{eq:S_mu}
    S_{\mu}(E,V)= \log \int_{H(\textbf{z};V)\leq E} d\textbf{z}
\end{equation}
is itself non-additive \cite{Gross02}. This is to stress that a
logarithmic structure does not ensure by itself additivity, even
within a standard ensemble, such as the microcanonical one. Both
Tsallis and microcanonical ensemble are potentially good in
describing non-additive systems, whereas the canonical is not.

In our view, it  is very important to stress that Eq.
(\ref{eq:dS}) holds independent of the number of degrees of
freedom of the system, i.e., the ensemble
(\ref{eq:Tsallis-ensemble}) is an \emph{exact orthode}
\cite{Campisi05}. In sum, the main result found here states that
the Tsallis ensemble provides a mechanical model of equilibrium
thermodynamics of non-additive and possibly small systems. We must
stress that exact orthodicity is not an exclusive feature of the
Tsallis ensemble, for example Boltzmann knew that the canonical
orthode is exact, and recently it has been stressed that the
microcanonical orthode is exact too \cite{Campisi05}. Therefore,
in principle all three ensembles are potentially good in
describing small dimensional systems. It is also worth stressing
that the result proved above holds for generic Hamiltonians,
\emph{without restriction to ideal non-interacting systems}.

It is useful to stress that, due to orthodicity, namely the fact
that standard thermodynamic relations exist among the quantities
$E, P, T, V, S$, one can construct the thermodynamic potentials as
usual by means of Legendre transforms. For example the free energy
will be given by:
\begin{equation}\label{}
    F = E - TS = E - T\log\mathcal{N}
\end{equation}
This is in complete agreement with the results of Ref.
\cite{Abe01} (compare with equation (29) therein), derived from
the requirement that the free energy should be expressed as a
function of the physical temperature $\beta^{-1}$. Here, the same
result has been obtained by adopting Boltzmann's general method,
namely by proving that the heat theorem holds within the Tsallis
ensemble. This approach is a very fundamental and unifying one.
Thanks to Boltzmann's concept of orthodicity, Tsallis ensemble has
been framed within the general theory of statistical ensembles
beside the microcanonical and the canonical ones. To better
appreciate this unifying feature, the reader is referred to
\cite{Campisi05}, where a proof, very similar indeed to the one
presented here, of the orthodicity of the microcanonical ensemble
is offered.

Let us now focus on the connection of this work with the standard
information theoretic approach where the leading role is played by
the Tsallis Entropy. It is quite simple to show that, for any
strictly monotonic $C^1$ function $g(x)$, the quantity
$\mathcal{N} g'(\mathcal{N}) \beta$ is an integrating factor for
$dE+PdV$ and the associated entropy is $S^{(g)} = g(\mathcal{N})$.
In fact, using Eq. (\ref{eq:deENdeE}) and (\ref{eq:deENdeV}) would
lead to
\begin{equation}\label{}
dS^{(g)}=\mathcal{N}g'(\mathcal{N}) \beta (dE+PdV).
\end{equation}
Adding the condition that the integrating factor be the average
kinetic energy, i.e., $\beta$, leads us to
$\mathcal{N}g'(\mathcal{N})=1$, which brings back to the
logarithmic entropy $S=log{\mathcal{N}}+const$. On the other hand,
adding the condition that the integrating factor equal the
Lagrange multiplier used in the Tsallis maximization procedure,
which in our notation would read $\beta_L = \beta
\mathcal{N}^{1/\alpha}$, leads to the Tsallis entropy form:
\begin{equation}\label{eq:Tsallis-entropy}
    S_q(\mathcal{N})=\alpha(\mathcal{N}^{1/\alpha}-1)+const= \ln_q
\mathcal{N}+const
\end{equation}
where the symbol $\ln_q$ indicates the q-deformed logarithm
\cite{Tsallis98}. Therefore
\begin{equation}\label{}
    dS_q = \beta_L \delta Q
\end{equation}
This fact has been acknowledged very recently also in
\cite{Abe06}. The present approach makes evident that the quantity
$\delta Q = dE+PdV$ admits infinitely many different integrating
factors associated with as many entropies and that the Tsallis
entropy is the one associated with the integrating factor
$\beta_L$. Therefore, adopting the Tsallis entropy, although the
integrating factor would not coincide with the average kinetic
energy, the thermodynamic relations would still keep holding.
Nonetheless the fact that the Lagrange multiplier $\beta_L$ does
not have such a straightforward physical interpretation as the
quantity $\beta$, poses some questions regarding the physical
interpretation of the associated entropy
(\ref{eq:Tsallis-entropy}) too. On the contrary, the fact that the
logarithmic entropy $S = \log \mathcal{N}$ is associated with the
\emph{physical temperature} i.e.,
\begin{equation}\label{}
    dS = \beta \delta Q
\end{equation}
makes it the ideal candidate to play the role of \emph{physical
entropy} or \emph{Clausius entropy} within the non-extensive
framework.

The fact that the present mechanical approach to non-extensive
thermodynamics does not lead to the Tsallis entropy, is not an
indication that an alternative information-theoretic approach,
consistent with the present, is lacking. Indeed it is quite easy
to prove that the logarithmic entropy (\ref{eq:S}), is equal to
the R\'{e}nyi entropy \cite{Renyi60}
\begin{equation}\label{eq:Rènyi-entropy}
\overline{S}[\pi]= \frac{\log \int d\textbf{z}
\pi(\textbf{z})^q}{1-q}
\end{equation}
of the ordinary distribution:
\begin{equation}\label{eq:ord-Tsallis-ensemble}
    \pi (\textbf{z};E,V) = \frac{\left[ 1 -
\frac{\beta}{\alpha}(H(\textbf{z};V)-E)\right]^{\alpha}}{\mathcal{N}(E,V)}
\end{equation}
where $E$ is the average energy evaluated over the normalized
distribution $\rho$ (\ref{eq:Tsallis-ensemble}). This fact follows
straightforwardly from Eq. (\ref{eq:n=N}), the identity
\begin{equation}\label{}
    \pi(\textbf{z};E,V)^q = \frac{\left[ 1 -
\frac{\beta}{\alpha}(H(\textbf{z};V)-E)\right]^{\alpha-1}}{\mathcal{N}(E,V)^q}
\end{equation}
and remembering the relation $\alpha = 1/(1-q)$.

Conversely it is easily seen that the distribution
(\ref{eq:ord-Tsallis-ensemble}), is obtained from the maximization
of the R\'{e}nyi entropy under the constraints:
\begin{eqnarray}\label{}
    \int d\textbf{z}\pi(\textbf{z}) &=& 1 \nonumber \\
   \frac{\int d\textbf{z}\pi(\textbf{z})^q H(\textbf{z})}{\int d\textbf{z}\pi(\textbf{z})^q}
   &=& E \nonumber
\end{eqnarray}
where the latter is evidently in accordance with the requirement
that the average energy should be evaluated over the normalized
distribution $\rho$. The maximization procedure shows that the
quantity $\beta$, namely the inverse physical entropy, coincides
with the Lagrange multiplier associated to the constraint on the
average energy in this case.

To conclude, we have shown that Tsallis ensemble is an exact
orthode, whose associated entropy is \emph{the logarithm of the
normalization function}, namely \emph{R\'{e}nyi entropy}. As
expected such entropy recovers canonical one and it is
non-additive. Further the result holds for small system (exact
orthodicity), where the interaction $\varphi(\textbf{q};V)$ is not
neglected. All this can be summarized by saying that Tsallis
ensemble provides a mechanical model of nonextensive
thermodynamics for small interacting Hamiltonian systems. The
present mechanical model based on the power-law ensemble leads to
R\'{e}nyi entropy, whose maximization, in turn, leads back to the
power-laws, thus preserving the the two-fold foundation (i.e.
mechanical and informational) typical of canonical ensemble.

\bibliography{thebibliography}% Produces the bibliography via BibTeX.

\end{document}